# First high-resolution spectroscopy of $^{201}$Hg$^+$ hyperfine structure: a sensitive probe of nuclear structure and the hyperfine anomaly

E.A. Burt, S. Taghavi-Larigani, and R.L. Tjoelker

*Jet Propulsion Laboratory, California Institute of Technology, Pasadena, CA*

Abstract

Using $^{201}$Hg$^+$ contained within a linear quadrupole rf ion trap that is part of an atomic clock, we report the first high-resolution spectroscopy on the hyperfine structure of $^{201}$Hg$^+$. We measure the absolute ground state hyperfine interval to be 29.954365821130(171)(62)(10) GHz, more than 8 orders of magnitude improvement over the previous measurement. The first error estimate in parentheses is the statistical error in the shifted line center measurement, the second is systematic uncertainty, and the third is calibration uncertainty in the hydrogen maser reference standard. By comparison to the already accurately known ground state hyperfine interval for $^{199}$Hg$^+$, we are able to derive a new value for the hyperfine anomaly in singly ionized mercury, $\Delta(S_{1/2}, ^{199}$Hg$^+, ^{201}$Hg$^+) = -0.0016257(5)$, now limited by knowledge of the nuclear magnetic moment ratio[1].

Keywords:

---

**INTRODUCTION**

Trapped mercury ions have long been employed in atomic clocks with applications including accurate primary frequency standards [1], commercial standards requiring continuous operation and high short-term stability [2], navigation [3, 4], and ultra-stable frequency flywheels [5-7]. Trapped ions are highly decoupled from their environment and mercury ions in particular have low sensitivity to external perturbations making it possible for clocks based on trapped mercury ions to reach exceptional levels of stability. To date all of the atomic clock work on trapped mercury ions has used the $^{199}$Hg$^+$ isotope. The other stable odd isotope (possessing ground state hyperfine structure) is $^{201}$Hg$^+$, which has seen virtually no application to atomic clocks. Very little is known about the sensitivity of its hyperfine clock transition to perturbing effects. We have begun a program to study $^{201}$Hg$^+$ and determine its viability for atomic clock applications. Part of our initial work is to accurately determine the $^{201}$Hg$^+$ clock transition frequency ($^2S_{1/2}$ $F = 1$, $m_F = 0$ to $^2S_{1/2}$ $F = 2$, $m_F = 0$), which up until now has only been known to 3 digits [8]. It is also possible to use accurate knowledge of the $^{201}$Hg$^+$ hyperfine clock transition frequency together with that of $^{199}$Hg$^+$ (already known) as a sensitive probe of nuclear structure.

Hyperfine structure in atoms is due to the interaction between the various nuclear magnetic and electric multi-pole moment of the nucleus and the field created there by orbiting electrons. A point nucleus can be assumed in a first-order calculation of hyperfine energies, but more accurate calculations must account for its finite size and the distribution of magnetization (the Bohr-Weisskopf effect [9]) and the distribution of electric charge (the Rosenthal-Breit effect [10]) contained within. In the case of an s-electron the magnetic contribution reduces to the Fermi contact interaction between the electronic spin magnetic moment and the nuclear magnetic moment, which depends on the fact that the s electron has a finite probability density at the origin. Thus atomic spectroscopy of hyperfine levels can be an excellent probe of nuclear structure. For a point nucleus, the ratio of the ground state hyperfine frequencies of two isotopes of the same atom is equal to the ratio of nuclear moments ($\mu$), nuclear spins (I) and total angular momenta (F):

$$\frac{\Delta f_1}{\Delta f_2} = \left(\frac{\mu_{I1}/I_1}{\mu_{I2}/I_2}\right)\frac{F_1}{F_2}. \quad (1)$$

The deviation from this expression due to the finite extent of the nucleus, referred to as the hyperfine anomaly, $\Delta$, can be expressed as:

$$\frac{\Delta f_1}{\Delta f_2} = (1+\Delta)\left(\frac{\mu_{I1}/I_1}{\mu_{I2}/I_2}\right)\frac{F_1}{F_2}. \quad (2)$$

In heavy elements $\Delta$ is on the order of $10^{-4}$ to $10^{-2}$. From a theoretical perspective, maximum information is obtained by measuring $\Delta$ in a chain of isotopes. This is particularly true in heavy elements such as francium [11] where the derived neutron radial distribution information can help constrain nuclear structure calculations essential to the understanding of parity non-conservation effects.

The first hyperfine anomaly measurement in neutral mercury performed by Eisinger [12] using the Knight shift gave a value of -0.0016(9). Subsequently, McDermott [13] derived a value for $\Delta$ of -0.001728(12), using a combination of optical excitation and varying magnetic field (the value was later corrected by the authors to -0.00164(3) – see comments in [14]). Building on Eisinger's method, Stager [14] improved the precision using that approach by a factor of 10 and found a value of -0.00175(9). Theoretical calculations are difficult to perform because they require precise knowledge of the valence electron wave functions but an estimate has been carried out that is in excellent agreement with experiment [15]. More recently Grandinetti used Electron Paramagnetic Resonance (EPR) to measure the anomaly in mercury ions held in a molecular lattice [16] and found values ranging from -0.0028(2) to +0.0056(4) dependent on temperature and molecular host. This large range includes the previous neutral mercury measurements, but is also consistent with values that are significantly different, raising the question as to whether the different electronic structure for the ion and neutral atom impacts the results. A later measurement in ions [17] gave an anomaly of -0.0034 at 10 K to -0.006 at 70 K, also consistent with the neutral measurements, and in partial agreement with the previous ionic measurement, but with a large range. Both cases had a temperature dependence that was unexpected.

Some of the most precise measurements of hyperfine anomalies have been performed in ion traps where the first-order Doppler shift can be eliminated in hyperfine spectra due to Lamb-Dicke confinement [18]. Measurements of the anomaly at the 1% level using trapped ions have been made in isotopes of Eu$^+$ [19] and Ba$^+$ [20]. In this paper we describe a first measurement of the anomaly in the stable 199 and 201 isotopes of Hg$^+$ using high-resolution spectroscopy on trapped Hg$^+$ ions.

To determine the anomaly, all measurements rely on knowledge of the nuclear magnetic moment ratio, now measured in mercury to be $\mu_{201}/\mu_{199} = -1.1074164(5)$ [21]. The uncertainty in the measurement of the mercury hyperfine anomaly is limited by the measurement of the hyperfine frequency ratio. In this paper we will describe an experiment that is part of our effort to develop a $^{201}$Hg$^+$ clock, which improves the knowledge of the 201/199 mercury frequency ratio by 8 orders of magnitude. We thereby obtain a new value for the mercury anomaly with a precision that is no longer limited by the frequency measurements.

**EXPERIMENTAL SETUP**

The measurements described here were performed on ions held within a linear Paul rf trap designed to work as a mercury ion atomic clock. The details of this apparatus have been described in [3]. Here we provide an overview of the aspects that are specific to this measurement.

*Apparatus Overview*

See Fig. 1 for a schematic of the ion trap region. A rf linear quadrupole ion trap is situated within an ultra-high vacuum titanium enclosure with a base pressure of approximately $10^{-7}$ Pa. The trap consists of 4 molybdenum rods evenly spaced on a circle of radius 6 mm. A trap drive of approximately 170 V at 1 MHz is applied to one pair of opposing rods with the same applied to the other pair 180 degrees out of phase. Neutral $^{201}$Hg is derived from enriched $^{201}$HgO (96% enriched; the natural abundance of $^{201}$Hg is 13.2%) inside an oven that operates at about 220 °C. Neutral $^{201}$Hg in the vicinity of the trap is bombarded by an electron pulse emitted from a heated LaB$_6$ cathode to create $^{201}$Hg$^+$. A significant portion of these ions have low enough energy to be captured by the trap, which has a well depth of about 3 eV. These are further cooled to near room temperature by a background helium buffer gas introduced into the vacuum enclosure from a heated quartz leak with a pressure of about $10^{-4}$ Pa. Typically about $10^7$ ions are loaded. Trapped ions are prepared in the desired initial internal state using a combination of light from a mercury discharge lamp at 194 nm and microwave radiation at approximately 29 GHz. The lamp consists of a fused silica envelope loaded with enriched $^{198}$Hg (92% enriched) in a 130 Pa buffer gas of argon. $^{198}$Hg has no ground state hyperfine structure and has a fortuitous electric dipole transition resonant only with the $^2S_{1/2}\ F=2$ to $^2P_{1/2}\ F=2$ transition of $^{201}$Hg$^+$ (see Fig. 2). With the discharge lamp switched to a dim state to avoid an AC Stark (light) shift, microwave interrogation consists of a Rabi pulse of microwave radiation applied at the first-order magnetic-field-insensitive $^2S_{1/2}\ F=1,\ m_F=0$ to $^2S_{1/2}\ F=2,\ m_F=0$ clock transition frequency. After microwave interrogation, the lamp is switched back to its bright state with light resonant only with the $^2S_{1/2}\ F=2$ to $^2P_{1/2}\ F=2$ optical transition at 194 nm. Ion fluorescence from this excitation at 194 nm is detected on a "solar-blind" photo-multiplier tube and indicates the degree to which the microwave interrogation frequency was on resonance with the clock transition. A pair of Helmholtz coils is used to generate the quantization axis magnetic field ("C-field") and the entire physics package, including optics, is enclosed in a 3-layer concentric cylindrical magnetic shield set.

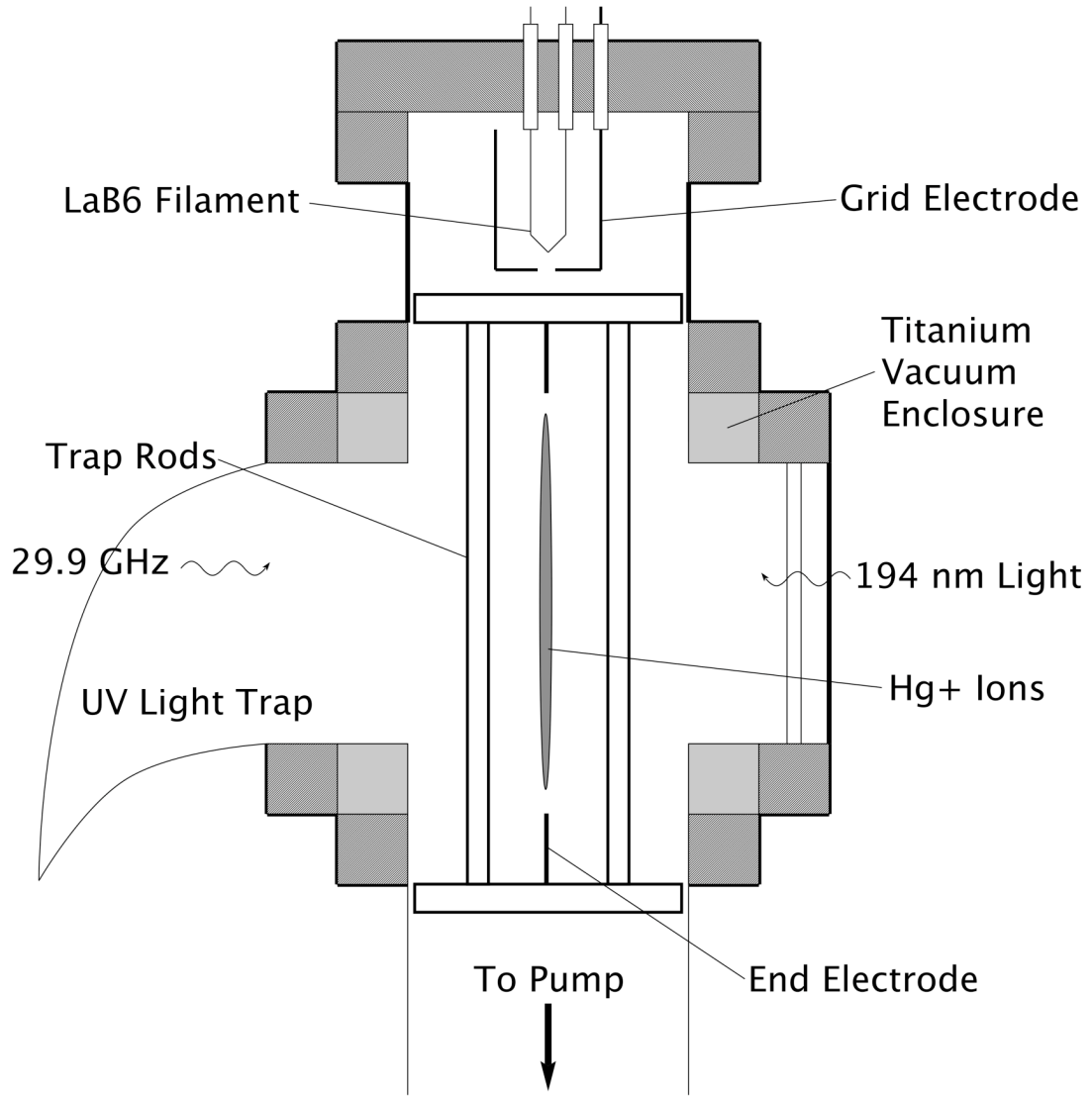

Figure 1. Mercury ion trap schematic diagram.

*State Preparation and Microwave Interrogation*

Before microwave interrogation of the clock transition can take place, we must prepare the $^{201}$Hg$^+$ ions in the $^2S_{1/2}$ $F = 1$, $m_F = 0$ clock state. In $^{199}$Hg+ this is straight forward because the F=0 hyperfine state has no Zeeman substructure: light resonant with the $F = 1$ ground state pumps ions into the dark $F = 0$, $m_F = 0$ clock state. However the level structure for $^{201}$Hg$^+$ (see Fig. 2) is more complicated and optical excitation of the $F = 2$ ground state alone will populate each of the 3 Zeeman sub-levels in the $F = 1$ ground state. To address this we employ a double resonance technique using both optical and microwave frequencies to clear out the $F = 1$, $m_F = \pm 1$ states. The $^{198}$Hg discharge lamp light source excites the $^{201}$Hg$^+$ $^2S_{1/2}$ $F = 2$ to $^2P_{1/2}$ $F = 2$ electric dipole transitions at 194 nm.

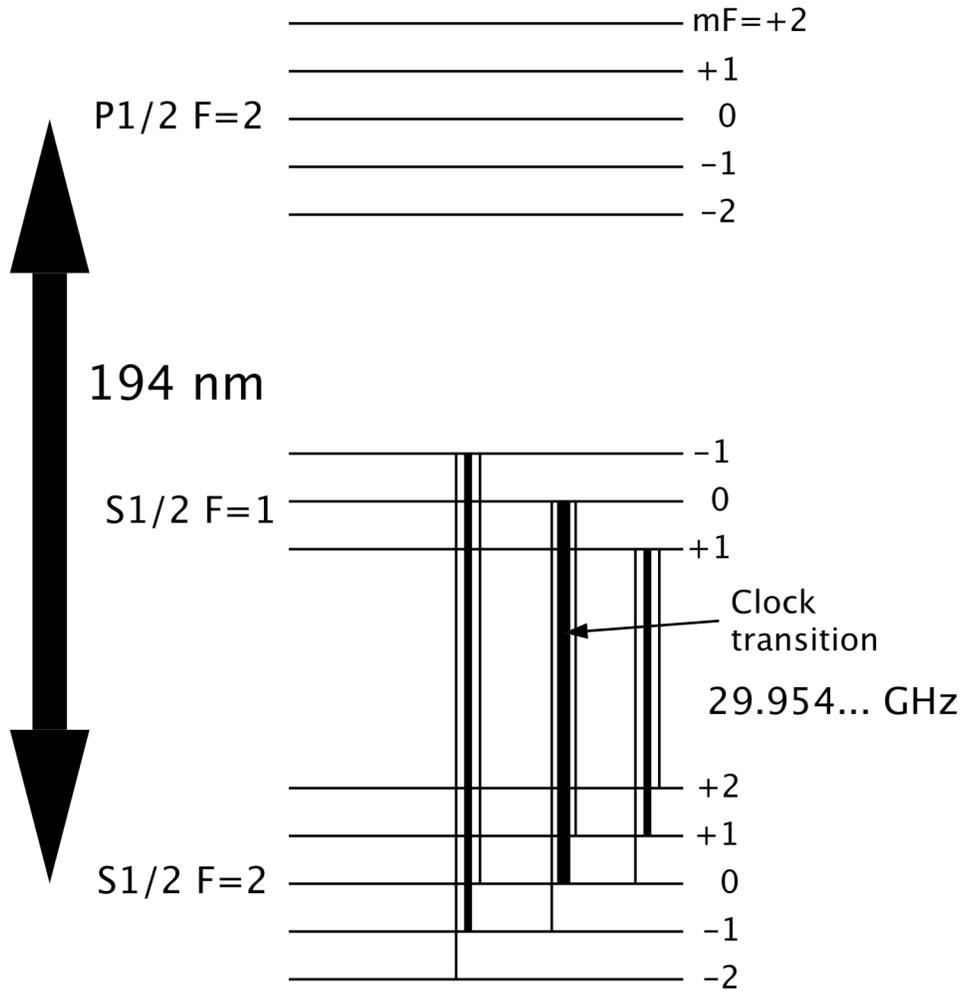

Figure 2. Energy level diagram for $^{201}$Hg$^+$ (not to scale). The S to P optical transition at 194 nm, shown as the thickest line, does not resolve the $m_F$ Zeeman structure. All possible microwave transitions between the $S_{1/2}$ hyperfine levels and their Zeeman sublevels are shown as thin lines with the lines used for microwave pumping ($m_F = -1$ to $-1$ and $m_F = +1$ to $+1$) shown slightly thicker. The clock transition ($m_F = 0$ to $0$) is shown as the thickest microwave line.

From the $^2P_{1/2}\ F = 2$ excited state, $^{201}$Hg$^+$ ions decay back into the $^2S_{1/2}\ F = 2$ or $^2S_{1/2}\ F = 1$ states. The $F = 1$ state has 3 Zeeman levels: $m_F = -1, 0$, and $+1$. At the same time that the lamp is on, we turn on microwaves resonant with the $^2S_{1/2}\ F = 1, m_F = \pm 1$ to $^2S_{1/2}\ F = 2, m_F = \pm 1$ transitions (highlighted in Fig. 2, these lines are several hundred kHz detuned from the $m_F = 0$ to $m_F = 0$ clock transition at a typical operating magnetic field of about 13.9 µT). These frequencies are generated as sidebands to a carrier near the clock transition, which is suppressed during this stage. In this arrangement ions will continue cycling until they decay into the dark $F = 1, m_F = 0$ ground state as desired.

After state preparation is complete, the microwave modulation and carrier suppression are turned off and a single microwave frequency at or near the clock transition for microwave interrogation is turned on. The microwave source is a high precision tunable synthesizer referenced to a hydrogen maser. The maser is itself calibrated to UTC(NIST) with an absolute fractional frequency uncertainty of less than $3 \times 10^{-13}$. Using a Ka-band microwave horn this microwave field is delivered to the trapped ions through the pyrex UV light trap shown in Fig. 1. The microwave polarization is parallel to the 13.9 µT applied magnetic field derived from the "C"-field Helmholtz coils so as to drive only $\Delta m = 0$ transitions. The Poynting vector of the interrogation microwaves, perpendicular to the magnetic field, is in the small radial direction of the linear trap, thus satisfying the condition for Lamb-Dicke confinement of the ions for this wavelength and enabling first-order Doppler-free spectroscopy. To the extent that the microwave frequency is resonant with the clock transition, the ions will be moved from $F = 1, m_F = 0$ to $F = 2, m_F = 0$ thereby making them once again resonant with the lamp. The lamp is again turned on and the subsequent ion fluorescence level is therefore a measure of the microwave frequency offset from resonance. This provides the mechanism to lock the microwave synthesizer (LO) to the ions and operate the system as a long-term stable atomic frequency standard. Or, with a very stable LO such as a hydrogen maser, it can be run open loop to perform high-resolution spectroscopy as described next.

# HIGH RESOLUTION SPECTROSCOPY OF $^{201}$Hg$^+$

Previously the $^2S_{1/2}\ F=1,\ m_F=0$ to $^2S_{1/2}\ F=2,\ m_F=0$ $^{201}$Hg$^+$ hyperfine clock transition frequency was measured at 29.94(15) GHz [8]. In our ion trap we searched for and found the clock transition within the error estimate of this value. The broad spectrum shown in Fig. 3 shows all hyperfine transitions possible in the ground state. Among the 9 transitions shown in Fig. 2, 2 pairs are degenerate, so only 7 lines are resolved in the spectrum shown here. These lines were highly power broadened to make them clearly visible on a broad-spectrum scan. Also evident in the spectrum are the radial ion motion (secular) sidebands at approximately ±35 kHz from each atomic line.

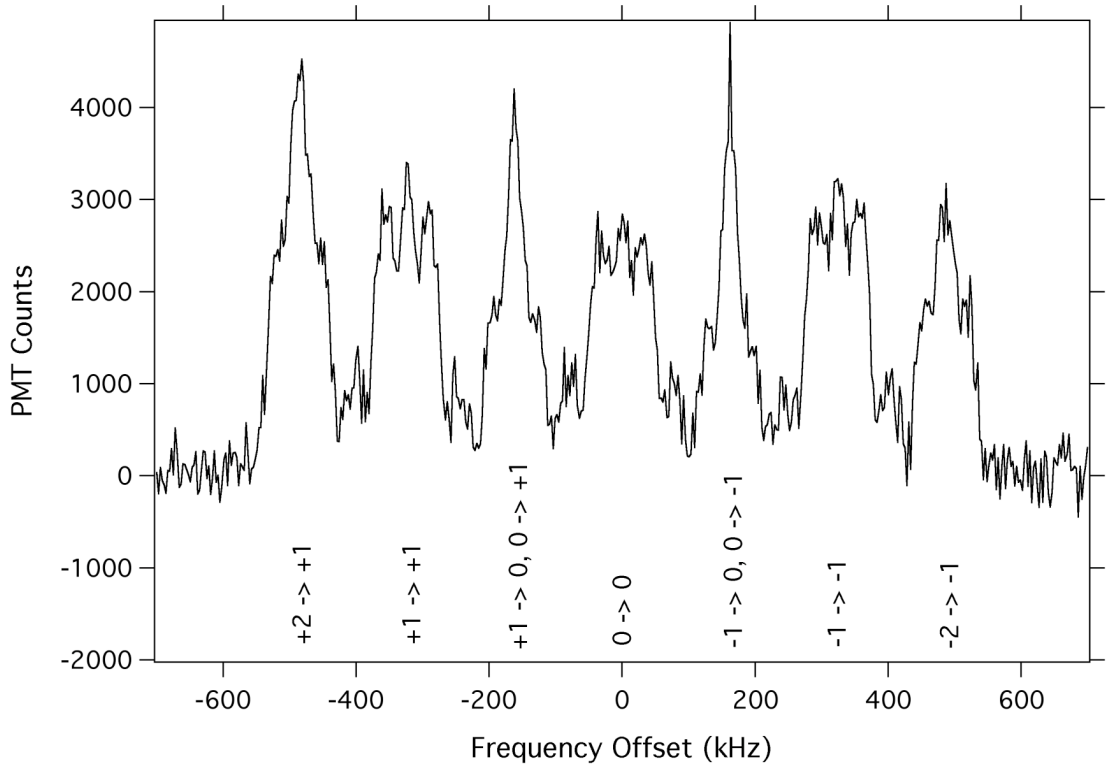

Figure 3. A wide-spectrum scan showing all hyperfine Zeeman transitions (11.5 µT field).

After optimization, a narrower spectrum at lower power with a better signal-to-noise ratio (Fig. 4), shows more detail. In addition to the central sharp clock transition are just resolved axial motion sidebands at approximately ±2 kHz, the radial motion sidebands at

±35 kHz, and the first Zeeman lines ($m_F = 0$ to $m_F = ±1$) at ±80 kHz (note that this spectrum was taken for a field of 11.5 µT).

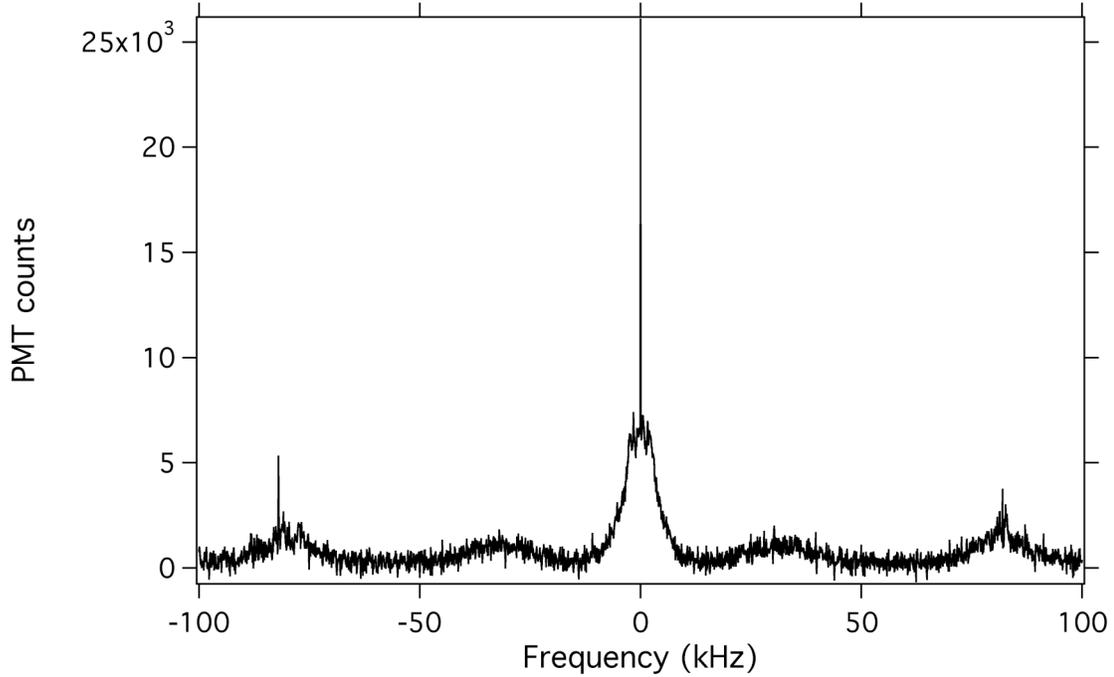

Figure 4. A narrower spectrum showing the central clock transition, motional sidebands and two of the Zeeman lines (11.5 µT field).

Once the hyperfine manifold was found a high-resolution scan of the clock transition at lower power was performed (see Fig. 5). Taking the average of the fitted central frequencies for several such spectra gives a (shifted) frequency of 29.954365823629(171) GHz at a field of about 13.9 µT. This data is obtained by applying a 1-second π-pulse to obtain a linewidth of approximately 1 Hz. The error estimate in the line center represents over 8 orders of magnitude improvement in resolution over the previous measurement.

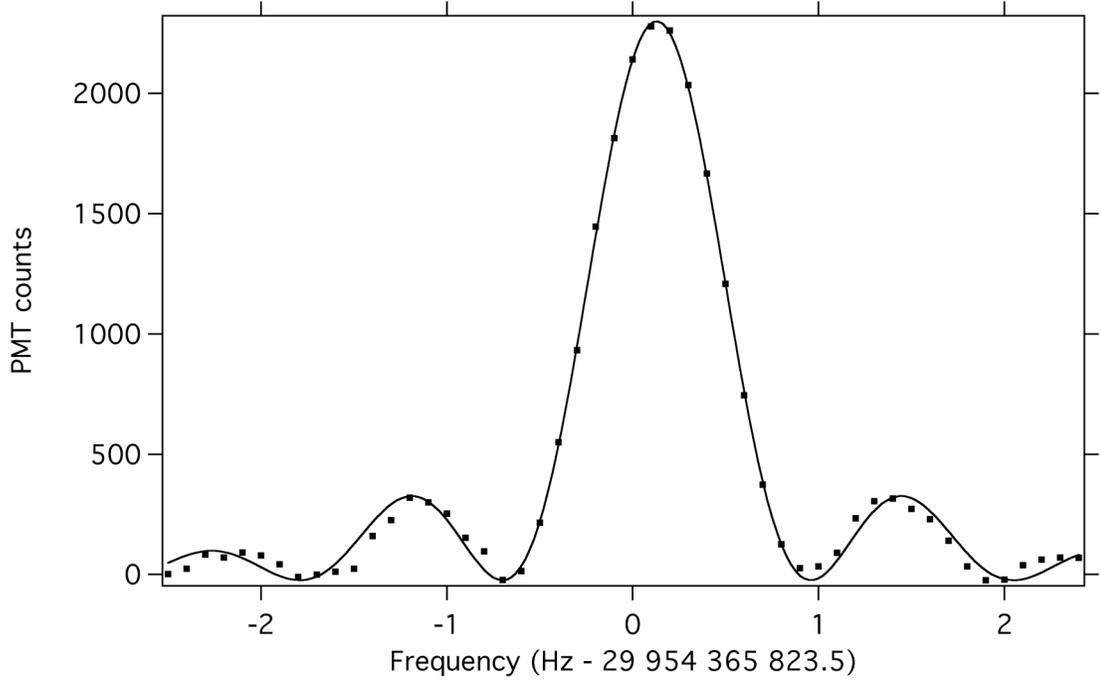

Figure 5. Shifted clock transition spectrum for a one-second Rabi pulse and a field of 13.9 µT. The data is fit to a Rabi lineshape. The slight discrepancy in the fit to the side lobe structure is due to a small residual broadening of the line, which is symmetric and has no effect on the center frequency.

## ERROR BUDGET AND ABSOLUTE $^{201}$Hg$^+$ CLOCK FREQUENCY

To determine the absolute frequency of the clock transition, and therefore the unshifted hyperfine interval, we estimate each of the largest frequency shifts present. These are the second-order Zeeman shift, second-order Doppler shift, shifts from collisions with background gases, and the AC Stark shift.

*Second-Order Zeeman Shift*

The first-order Zeeman shift is suppressed in the clock transition, however the second-order Zeeman shift is still by far the largest systematic Effect. We use the ions themselves as a magnetometer by probing the first-order field sensitive transition, $^2S_{1/2}\ F = 1,\ m_F = -1$ to $^2S_{1/2}\ F = 2,\ m_F = -1$, which is also first-order Doppler free (since $\Delta m = 0$ for this transition the Poynting vector for the required microwave radiation is parallel to the small radial direction of the trap, as with the clock transition, and the ions are Lamb-Dicke confined for both this field-sensitive transition and the clock transition). Excellent precision is possible as shown in Fig. 6.

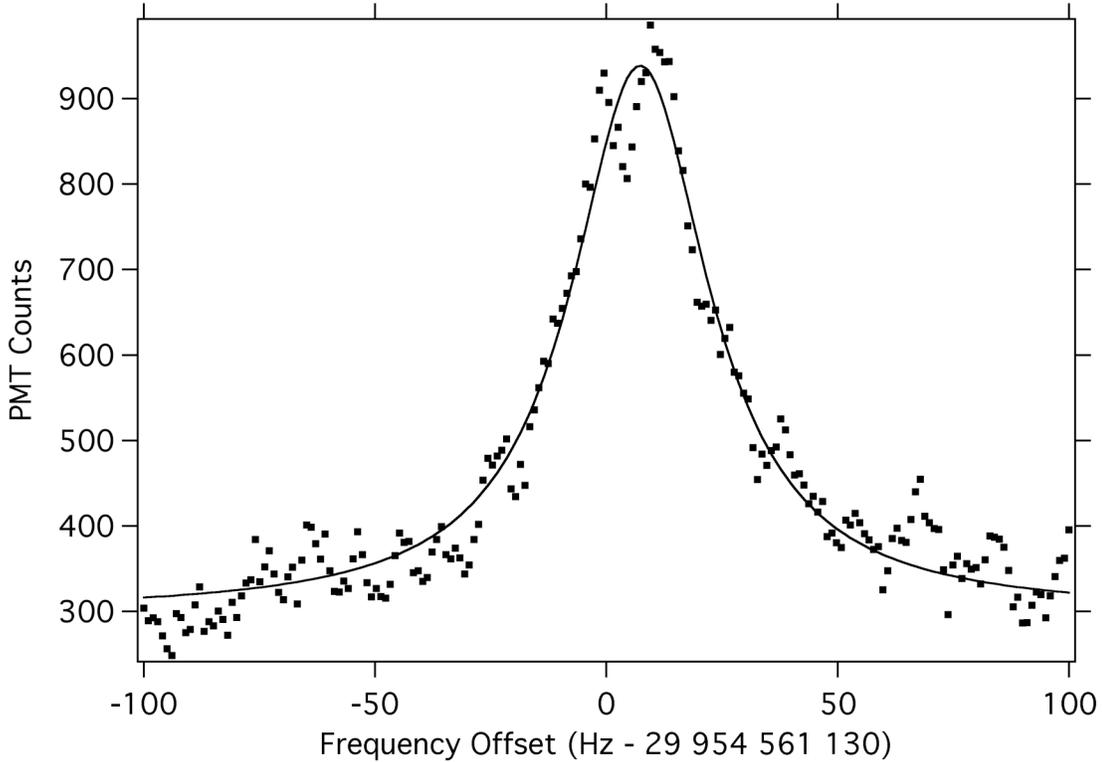

Figure 6. High resolution spectroscopy of the $m_F = -1$ to $m_F = -1$ Zeeman line. The spectrum is broadened by magnetic field noise and higher microwave power used to boost signal.

A fit gives a field-sensitive line center of 29.954561137(30) GHz, or approximately 195 kHz from the clock transition at the current field setting. The relatively large error estimate reflects the 50 ppm/°C stability of the C-field current source and the 3-4°C daily temperature fluctuations that we experience in the development area of our lab, which doesn't have precise temperature control. The Breit-Rabi formula gives the total magnetic frequency shift to second-order for both the field-sensitive and field-insensitive lines:

$$f = -f_0\left(1 + m_F x + x^2\right)^{1/2}, \text{ with } x = (g_J + g_I)\mu_B B_0/hf_0, \tag{1}$$

where $g_J = 2.0031745(74)$ [22] and $g_I = -2.03406(12)\times 10^{-4}$ [23] are the electronic and nuclear g factors respectively, $\mu_B$ is the Bohr magneton, $B_0$ is the applied field and $f_0$ is the un-shifted absolute clock transition frequency. The value for $g_J$ was measured in $^{198}$Hg$^+$. In [22] the authors state that differences in $g_J$ between mercury isotopes should be much smaller than the error estimate for the measured value in $^{198}$Hg$^+$. Therefore we take the value for $^{198}$Hg$^+$ to be representative of the value for $^{201}$Hg$^+$. Inserting our measured values for the shifted clock and field-sensitive frequencies gives a second-order Zeeman shift of +2.5471(5) Hz.

*Second-Order Doppler Shift*

First-order Doppler shift effects are suppressed by Lamb-Dicke confinement, but there is a second-order Doppler shift because the time-average of the ion velocity squared as it executes periodic motion in the trap is non-zero. There are two components of this effect: one due to ion temperature and the other due to rf trap drive induced micromotion. We have not measured the temperature of $^{201}$Hg$^+$ ions yet, but have measured the ion temperature of $^{199}$Hg$^+$ ions in the same trap with the same buffer gas pressure to be approximately 650 K [24]. The temperature-dependent second order Doppler shift is given by:

$$\Delta f = -f_0 \frac{3k_B T}{2mc^2}, \tag{3}$$

where $k_B$ is Boltzmann's constant, $T$ is the temperature, m is the mass of the ion and $c$ is the speed of light. For $T = 650$ K, the shift is -13 mHz. Since the mass difference between the two isotopes is only 1%, we expect to see the same frequency shift (at our overall current level of precision) for the two isotopes. However since we have not measured the ion temperature directly for $^{201}$Hg$^+$ we assign this entire shift to the uncertainty in this effect.

As the number of trapped ions changes, Coulomb repulsion results in a different ion cloud size so that, on average, ions will sample a different trap drive rf amplitude. This leads to a different average micromotion velocity and thus an ion-number-dependent second-order Doppler shift. This effect has been measured in this trap extensively for $^{199}$Hg$^+$ ions [25]. The total magnitude is approximately -40 mHz for a full trap. Numerical integration of the micromotion-induced second-order Doppler shift over the volume of the trap weighted by the numerically-determined ion density predicts that the difference in sensitivity to this effect between the two isotopes ($^{199}$Hg$^+$ and $^{201}$Hg$^+$) is less than 1%. As before we assign the magnitude of the entire shift (40 mHz) to the uncertainty in this effect.

*Pressure Shifts*

Vetter has derived an approximate expression for the fractional frequency shift in the hyperfine transition of alkali-like ions due to collisions with a background gas [26]. For different isotopes of the same atom, the theoretical expression suggests that the frequency shift for a given background gas should scale as the ground state valence electron orbital radius. Since the nuclear charge is the same for the two isotopes considered here and the nuclear mass differs by only 1%, the expectation value of the ground state valence electron radius should differ by less than 1%. Thus, for our present level of precision, we use the pressure shift constants measured for $^{199}$Hg$^+$ [27]. Precise measurements of the pressure shift constants for $^{201}$Hg$^+$ will be the subject of future work. Our primary background gas is helium with a fractional frequency shift constant of $+1.6 \times 10^{-10}$/Pa. The helium pressure was $7.7(4) \times 10^{-6}$ Pa, so that the entire fractional shift is $+1.2(1) \times 10^{-13}$ or 3.7(2) mHz. All other background gas pressures are less than $1.3 \times 10^{-7}$ Pa. Of these, the one with the largest shift constant is methane (CH$_4$) at $-2.7 \times 10^{-7}$/Pa. In the worst case where the entire base pressure of $1.3 \times 10^{-7}$ Pa is due to methane, the associated fractional shift would be $-3.6(3.6) \times 10^{-14}$ or -1.1(1.1) mHz. All other pressure shifts are several orders of magnitude smaller and are not included.

*AC Stark Shift*

To avoid an AC Stark shift ("light shift") from the discharge lamp, the lamp is switched to a dim mode during interrogation of the clock transition. The lamp intensity for the dim state is about 100 times lower than that of the bright state used for optical pumping and atomic state readout. Accounting for lamp spectral characteristics and Doppler broadening of the optical dipole transition, we estimate an AC Stark shift for the clock transition during normal operation on the order of 1 mHz. By measuring the clock frequency with the lamp in two different dim states differing in intensity by a factor of 20 and extrapolating to zero-intensity, we are able to place a bound on the AC Stark shift during normal operation of 0.002(45) Hz. All other systematic effects in general are several orders of magnitude smaller than those considered here.

*Derived Accuracy*

Finally, all frequencies described here are values indicated on the precision tunable synthesizer referenced to a hydrogen maser that is NIST-calibrated to within $3 \times 10^{-13}$ (this calibration is monitored and maintained using GPS time transfer). Therefore the "derived" uncertainty in all of the hyperfine frequency measurements is 10 mHz.

Table 1 summarizes these results. The total systematic shift is +2.499(62) Hz. Applying this to the measured clock transition frequency of 29.954365823629(171) GHz gives an un-shifted absolute clock transition frequency of:

$$f_0 = 29.954365821130(171)(62)(10) \text{ GHz}, \qquad (4)$$

where the first error estimate is the statistical uncertainty, the second is the total systematic uncertainty derived from the values shown in Table 1 combined in quadrature (without the derived accuracy), and the third is the derived accuracy.

Table 1. Largest systematic shifts of the $^{201}$Hg$^+$ hyperfine clock frequency. Note that the total error estimate for the systematic effects does not include statistical uncertainty in the measurement of the shifted line center and derived accuracy. These two are called out separately.

| Effect | Shift (Hz) | Error Estimate (Hz) |
|---|---|---|
| Second-order Zeeman | +2.5471 | 0.0005 |
| Second-order Doppler (T) | -0.013 | 0.013 |
| Second-order Doppler (N) | -0.04 | 0.04 |
| Pressure shift (He) | +0.0037 | 0.0006 |
| Pressure shift (CH$_4$) | -0.0011 | 0.0011 |
| AC Stark shift | +0.002 | 0.045 |
| *Total Shift* | +2.499 | 0.062 |
| Statistical Uncertainty | - | 0.171 |
| Derived Accuracy | - | 0.01 |

**THE HYPERFINE ANOMALY IN Hg$^+$**

Using the present $^{201}$Hg$^+$ clock frequency measurement (Eq. 4) and the NIST value for the $^{199}$Hg$^+$ clock transition frequency of 40.50734799684159(41) GHz [1], we obtain the ratio (for simplicity we include in the error estimate for the ratio only the largest uncertainties, which dominate the others, in each of the frequency measurements):

$$\frac{f_{201}}{f_{199}} = -0.739479805577(5). \tag{5}$$

Note that the ratio is negative because the hyperfine structure of $^{201}$Hg$^+$ is inverted (see Fig. 2). All states involved in both transitions are composed of s electrons with total electronic spin J = ½ and so there is no contribution to this hyperfine structure from an electric quadrupole moment in the nucleus. We will also assume that higher order magnetic octopole and electric hexadecapole contributions are negligible so that only the magnetic dipole moment of the nucleus need be considered here. From the ratio in Eq. 5 we derive a hyperfine anomaly in the $S_{1/2}$ state of,

$$\Delta\left(S_{1/2},^{199}Hg^+,^{201}Hg^+\right) = -0.0016257(5), \tag{6}$$

where the error estimate is determined by the uncertainty in the measurement of $\mu_{201}/\mu_{199}$. This new value for the anomaly is in statistical agreement with the previous neutral measurement by McDermott [13], but in slight disagreement with that of Stager [14] with almost two orders of magnitude better precision (now limited by the magnetic moment ratio instead of the frequency ratio). Our measurement also agrees with neutral mercury theory, and falls within the range of previous ionic measurements. This result indicates that the shielding due to the second electron in the neutral atom has at most only a small effect on the anomaly at this precision level. We hope that this work will stimulate further studies to improve the theoretical estimate of the anomaly.

The 199 and 201 isotopes of mercury are the only stable mercury isotopes possessing ground state hyperfine structure. However, the method used here to measure the hyperfine frequency and to derive the hyperfine anomaly can be performed with similar precision in only hours of integration time. Thus, if a radioactive source of mercury were available, it would be feasible to extend the measurement to a longer chain including unstable isotopes, with benefits to nuclear physics as outlined in [28].

**CONCLUSION**

As part of a program to develop a $^{201}Hg^+$ trapped ion atomic clock, we have performed a new measurement of the $S_{1/2}$ $F = 1$, $m_F = 0$ to $S_{1/2}$ $F = 2$, $m_F = 0$ hyperfine clock transition in $^{201}Hg^+$ ions and found a frequency of 29.954365821130(171)(62)(10) GHz. This new value is 8 orders of magnitude more accurate than the previous best measurement. We have combined this value with the hyperfine clock transition frequency in $^{199}$Hg$^+$ to give a new value for the hyperfine anomaly in mercury of -0.0016257(5). The new value is in agreement with previous measurements in neutral mercury suggesting that shielding of this effect from the extra valence electron in neutral atoms is not significant within the combined error estimate of neutral and ionic values.

**ACKNOWLEDGEMENT**

The authors gratefully acknowledge helpful discussions with Z.-T. Lu on nuclear structure and to W. Itano on the hyperfine anomaly.

20